\documentclass{nature}
\usepackage{verbatim}
\usepackage{amsmath}
\usepackage{xcolor}
\usepackage{graphicx}
\makeatletter
\let\saved@includegraphics\includegraphics
\AtBeginDocument{\let\includegraphics\saved@includegraphics}
\usepackage{caption}
\renewenvironment*{figure}{\@float{figure}}{\end@float}
\makeatother

\title{Switching dynamics of dark-pulse Kerr comb states in optical microresonators}

\author{Elham Nazemosadat$^1$, Attila F\"{u}l\"{o}p$^{1,\text{A}}$, \'{O}skar B. Helgason$^1$, 
Pei-Hsun Wang$^{2,\text{B}}$, Yi Xuan$^{3,\text{C}}$, Dan E. Leaird$^2$, Minghao Qi$^{2,3}$, Enrique Silvestre$^4$, Andrew M. Weiner$^{2,3}$ \& Victor Torres-Company$^1$}

\begin{document}

\maketitle

\begin{affiliations}
 \item Department of Microtechnology and Nanoscience (MC2), Chalmers University of Technology, SE-41296 G\"{o}teborg, Sweden.
 
 \item School of Electrical and Computer Engineering, Purdue University, West Lafayette, IN 47907-2035, USA.
 
 \item Birck Nanotechnology Center, Purdue University, West
Lafayette, IN 47907-2035, USA.

 \item Department of Optics-ICMUV, University of Valencia, 46100 Burjassot, Valencia, Spain.
 
 $^\text{A}$ Now at OptiGOT AB, SE-41133 G\"{o}teborg, Sweden.
 
 $^\text{B}$ Now at Department of Optics and Photonics, National Central University, Taoyuan City 32001, Taiwan.
 
 $^\text{C}$ Now also at Department of Surgery, Indiana Center for Regenerative Medicine and Engineering, Indiana University School of Medicine, Indianapolis, IN 46202, USA.
\end{affiliations}

\begin{abstract}
Dissipative Kerr solitons are localized structures that exist in optical microresonators. They lead to the formation of microcombs –-- chip-scale frequency combs that could facilitate precision frequency synthesis and metrology by capitalizing on advances in silicon photonics. Previous demonstrations have mainly focused on anomalous dispersion microresonators. Notwithstanding, localized structures also exist in the normal dispersion regime in the form of circulating dark pulses, but their physical dynamics is far from being understood. Here, we report the discovery of reversible switching between coherent dark-pulse Kerr combs, whereby distinct states can be accessed deterministically. Furthermore, we reveal that the formation of dark-pulse Kerr combs is associated with the appearance of a new resonance, a feature that has never been observed for dark-pulses and is ascribed to soliton behavior. These results contribute to understanding the nonlinear physics in few-mode microresonators and provide insight into the generation of microcombs with high conversion efficiency.
\end{abstract}
Dissipative solitons are self-enforcing, stationary structures that exist in diverse nonlinear dissipative systems subject to an external pump of energy~\cite{Akhmediev}. The recent discovery of temporal dissipative solitons in optical cavities displaying Kerr nonlinearity~\cite{Leo2010,Herr2013} (from now on dissipative Kerr solitons or DKS) has facilitated the investigation of their rich dynamics~\cite{KippenbergScience,Godey,Coen:OL13,Matsko:12,Bao16,Yu2017,Lucas2017,Cole2017,Wang:18,Karpov2019,Guo2016,PASQUAZI2018}. DKS rely on balancing the inherent cavity dispersion with the corresponding Kerr nonlinear phase shift induced by the soliton, while the dissipative nature of the microresonator is offset by supplying the cavity with the energy from a pump laser. DKS are just one particular solution of the complex spatio-temporal landscape in nonlinear Kerr cavities~\cite{Godey,Coen:OL13}. The same microresonator can also display chaos, breathing dynamics~\cite{Matsko:12,Bao16,Yu2017,Lucas2017}, soliton crystals~\cite{Cole2017,Wang:18} and transitions between some of these states~\cite{Karpov2019}. The single soliton regime can be accessed deterministically by decreasing the number of cavity solitons while properly tuning the pump laser over the resonance~\cite{Guo2016}. Mapping this complexity is not only of fundamental interest, but important for the design and operation of stable, ultra-broadband coherent Kerr frequency combs in high-Q microresonators (microcombs)~\cite{KippenbergScience,PASQUAZI2018}, which have potential applications in multiple fields, ranging from optical clocks to coherent communications~\cite{Marin-Palomo2017,Suh,Liang2015,Kippenberg:LIDAR,Vahala:LIDAR,Vahala:Vernier,Spencer2018,Obrzud2019,Vahala:calibration,Drake2019,Xue:RF}.

DKS require the optical microresonator to display anomalous dispersion~\cite{note1} at the pump wavelength. 
Interestingly, other stationary structures such as  ultrashort optical pulses~\cite{HuangPRL} or dark-pulse Kerr combs~\cite{Xue2015} can be found in high-Q microresonators operating in the normal dispersion regime (i.e. decreasing free spectral range (FSR) with optical frequency). As the name implies, the time-domain waveform of a dark-pulse Kerr comb corresponds to a localized dark-pulse structure, where low intensity oscillations are embedded in a high intensity background. These pulses can be interpreted as two stably interlocked switching waves, connecting the upper and lower homogeneous steady-state solutions of the bi-stability curve in Kerr microresonators~\cite{Parra-Rivas:16}. These localized waveforms also exhibit breathing dynamics~\cite{Bao:2018} and have intriguing connections to sneaker waves found in hydrodynamics, called flaticons~\cite{Varlot:13} and platicons~\cite{Lobanov:15} in optics. In comparison to DKS, the physics of dark-pulse Kerr combs is less understood due in part to a complex interplay between multiple modes and thermal dynamics in the cavity, but these microcombs are more efficient in converting the pump power into useful comb light~\cite{Xue:2017} –-- an aspect that is particularly promising for coherent optical communications~\cite{Attila2018,Helgason:19}. Some key questions remain unanswered, such as what the pathway to their generation is, starting from a continuous-wave (CW) waveform, and whether this transition is accompanied by similar switching dynamics to what has been observed in DKS.

In this work, we report deterministic switching between dark-pulse Kerr comb states, where each state is uniquely ascribed to a number of low intensity oscillation periods. This number can be deterministically controlled and increased or decreased one at a time, unraveling an overlooked dependence with the pump laser detuning parameter for dark-pulse Kerr combs. Our results are in excellent agreement with numerical simulations that take naturally into account the linear coupling between the dominant transverse modes of the microresonator. Strikingly, we find that the formation of dark-pulse Kerr combs is also accompanied by the appearance of an extra resonance, in compelling similarity to the behavior reported for DKS~\cite{Guo2016} and perfect soliton crystals~\cite{Karpov2019}. In contrast, however, {\color{black}our measurements reveal that in dark-pulse Kerr combs, the pump is effectively blue-detuned with respect to the cavity resonance that is Kerr shifted due to the high power CW background of the dark-pulse.}
\section*{Results}
\begin{figure}[tb]
\centering
\includegraphics[width=.8\linewidth]{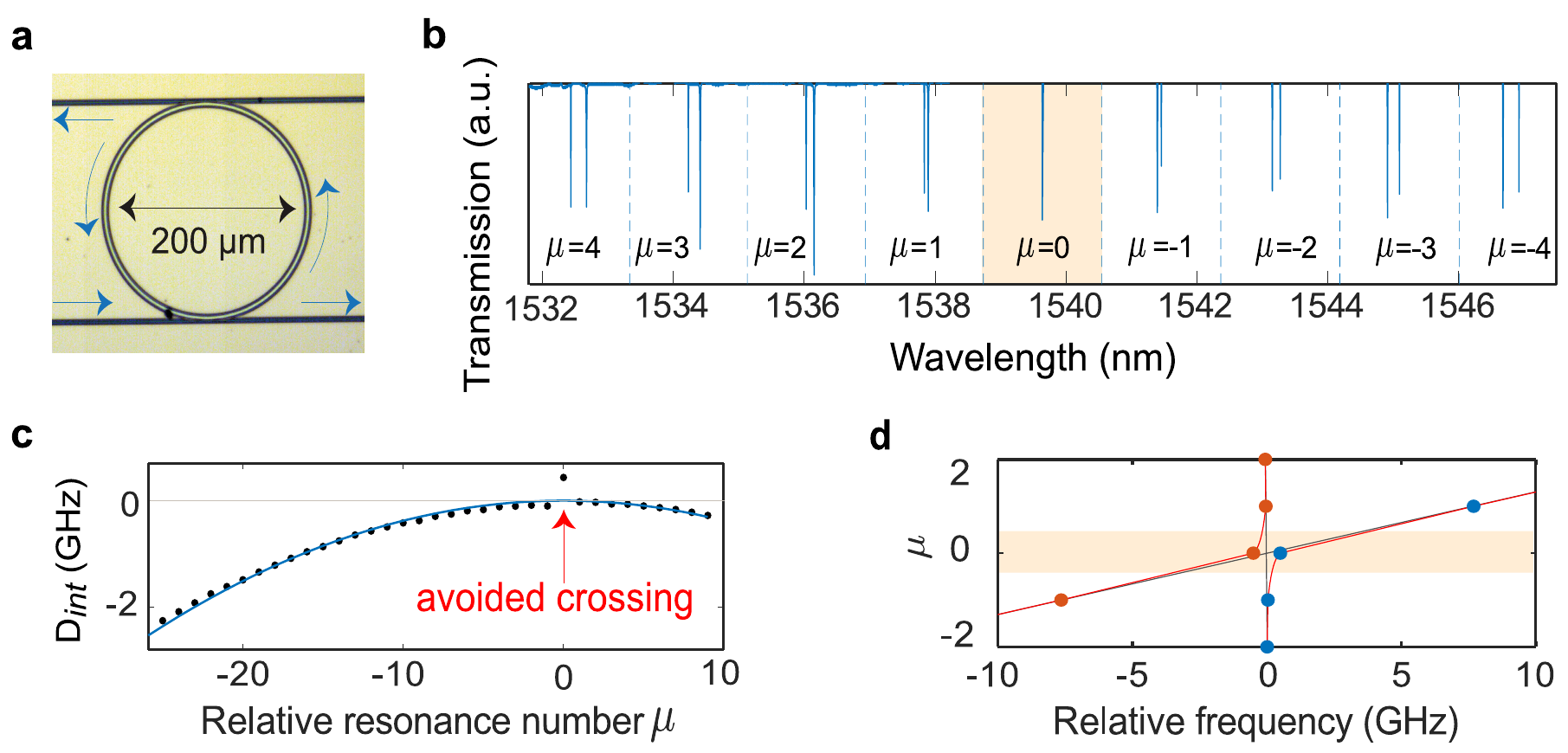}
\caption{Linear characterization of the multi-mode silicon nitride microresonator. \textbf{a}, Microscope image of the silicon nitride microresonator. 
\textbf{b}, Measured transmission scan of the microresonator, where two clear transverse modes appear. \textbf{c}, Integrated dispersion, i.e. frequency deviation of the resonance locations of the main mode (black dots), $\omega_{\mu}$, with respect to an ideal grid, $D_{int}=(\omega_{\mu}-\omega_0-2\pi D_1 \mu)/2\pi$, where  $\mu$ is the mode number and $D_1$ the free spectral range. The main mode displays normal dispersion, while at $\mu=0$ the dispersion changes locally to anomalous due to the linear coupling between two transverse modes in the resonator. 
\textbf{d}, Linear mode coupling effect. The transmission spectrum is divided into blocks with a spacing difference of 1 FSR (dotted lines in \textbf{b)} to plot this diagram and calculate the linear coupling strength. Mode coupling shifts the resonances apart from each other (avoided modal crossing). The group velocity dispersion, mean intrinsic Q and linear coupling parameters are $\beta_2=139~\rm{ps^2/km}$, 1.6 million and $\kappa=22.7~\rm{m^{-1}}$, respectively.
}
\label{fig:ring}
\end{figure}

\subsection{Microresonator characterization in the linear regime.}
A silicon nitride microresonator with a designed cross-section of 2~$\mu$m (width) $\times$ 600 nm (height) is used in our experiments (Fig.~\ref{fig:ring}a). The particular modes of interest are $\text{TE}_1$ and $\text{TE}_2$, which exhibit normal dispersion within the C~band~\cite{Wang:13}. The fabrication process for this design has been described elsewhere~\cite{Xuan:16}. The ring features a radius of 100 $\mu$m, corresponding to an FSR of around 229 GHz for the main mode used for comb generation, with a measured mean intrinsic Q-factor of around 1.6 million. A tunable external-cavity pump laser with sub-10 kHz linewidth is used for pumping the microresonator. {\color{black}It is calibrated using a fiber Mach-Zehnder interferometer~\cite{HuangPRL}. To characterize the ring, the pump is scanned over the C-band to find the resonance locations of the two linearly coupled transverse modes in the microresonator.} 
The measured transmission scan shown in Fig.~\ref{fig:ring}b displays an avoided mode crossing around 1540 nm, which is due to linear mode coupling, and results into a local change of dispersion (Fig.~\ref{fig:ring}c and~\ref{fig:ring}d), thus facilitating phase matching for parametric oscillation from a CW pump~\cite{Liu:14,Xue2015}. 
For comb generation, the set of two hybridized resonances resulting from the avoided mode-crossing is pumped from the blue side. The same microresonator, pumped in the same way, has been previously used to generate a mode-locked Kerr comb, with evidence of dark-pulses circulating in the cavity~\cite{Attila2018}.

\subsection{Switching dynamics of dark-pulse comb states.}
\begin{figure}
\centering
\includegraphics[width=\linewidth]{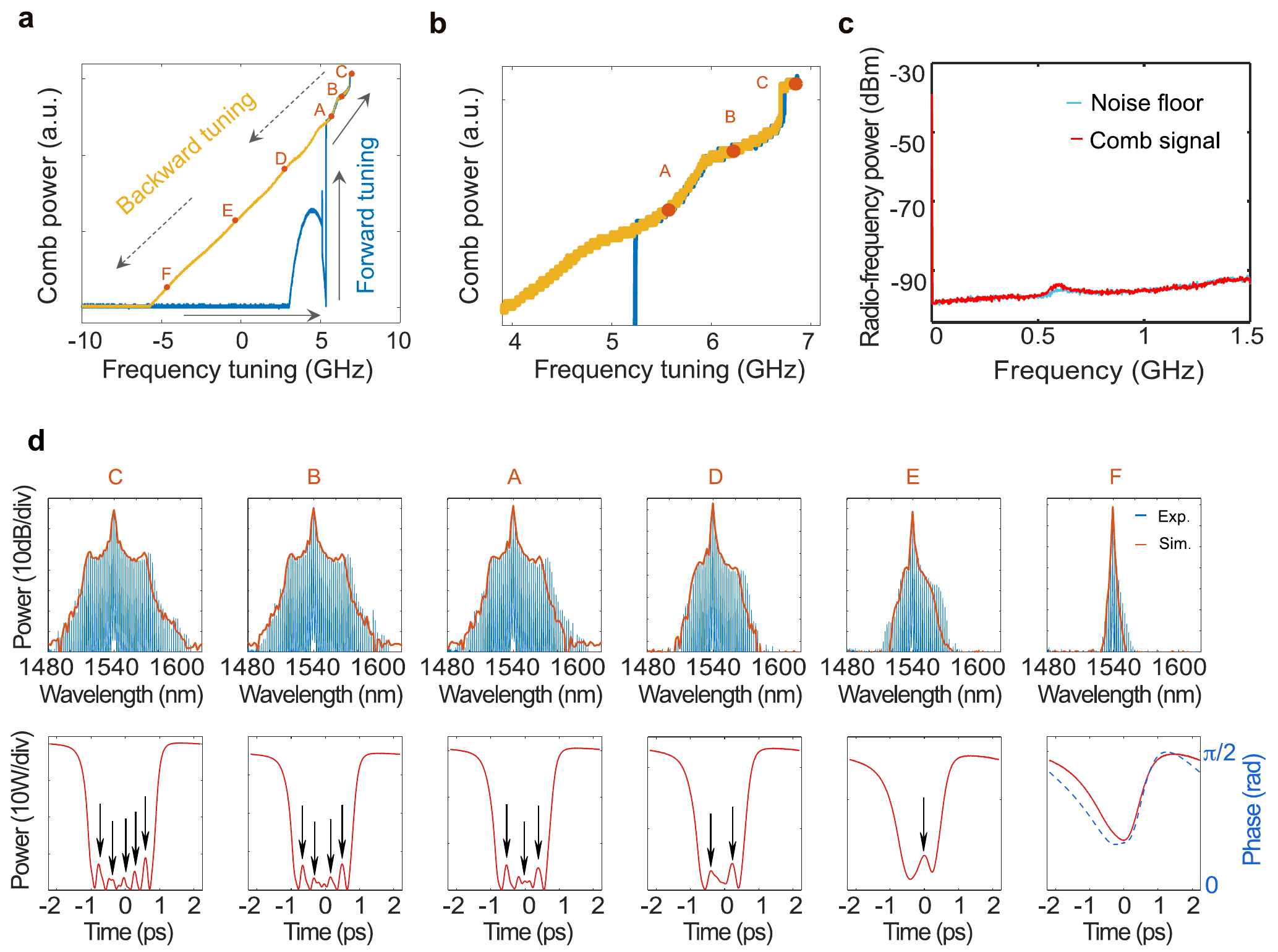}
\caption{Deterministic switching of dark-pulses. \textbf{a}, Measured comb power when the pump is forward (blue) and backward (yellow) tuned. \textbf{b}, Step-like patterns are observed as the pump is tuned, indicating switching between dark-pulse comb states. The pump is first forward tuned from the blue side towards the red, where the combs A-C are generated. After accessing the comb state C, the pump is tuned back towards the blue side. In the backward tuning, switching over a broader detuning range is observed. \textbf{c}, The RF spectrum (red) and the noise floor (blue) of the generated comb, confirming that the comb is operating in a low-noise state. \textbf{d}, The blue frequency lines are the comb spectra measured at different pump detunings, corresponding to the comb states marked in \textbf{a}. The simulated comb envelope of each state is shown in red, with the corresponding time-domain waveforms underneath. The arrows point the number of low intensity oscillations. For state F, the phase of the pulse is also shown.
}
\label{fig:Exp_Switching}
\end{figure}

We pump the resonance that experiences the stronger linear coupling, indicated by the red arrow in Fig.~\ref{fig:ring}c. As the pump is tuned over the resonance from the thermally stable blue side towards the red~\cite{Carmon:04,Herr2013} (forward tuning), it enters the dark-pulse existence range and a coherent dark-pulse Kerr comb is generated (see Methods). The comb power increases in a step-like manner {\color{black}as seen in the top-most part of Figs.~\ref{fig:Exp_Switching}a,b (positions A$\rightarrow$ B $\rightarrow$ C).} This behavior is similar to previous observations made by Xue et al.~\cite{Xue2015}. Here, we elucidate that these steps correspond to a transition between different dark-pulse comb states. Each step corresponds to a coherent comb state, indicated by a low amplitude noise as shown in Fig.~\ref{fig:Exp_Switching}c, that can be accessed sequentially. 
Keeping the pump power fixed, the comb power measurements in the forward pump tuning are repeated 100 times and step-like patterns that are almost identical to those shown in Fig.~\ref{fig:Exp_Switching}a are measured. 
The dynamics reveal that at the used power level, the comb does not go over a chaotic state, making the comb generation process repeatable and deterministic. The comb found in state A achieves a conversion efficiency of around 25\%, {\color{black} where the conversion efficiency is defined as the output power in the comb lines (excluding the pump) divided by the input pump power.}

To get a better insight into the physics of comb generation in the normal dispersion regime, we simulate our experimental findings using an Ikeda map \cite{IKEDA,Torres-Company:14,Hansson:15}, modified such that the linear mode coupling in the cavity is taken into account (see Methods for implementation details). 
The parameters used in the simulations are extracted from the transmission scan measurements (Fig.~\ref{fig:ring}b). The simulated comb spectra for various pump detunings and their corresponding simulated intracavity waveforms are shown in red in Fig.~\ref{fig:Exp_Switching}d. The simulated time-domain waveforms reveal that in the forward pump tuning, with each step in the comb power, one low intensity oscillation appears at the center of the dark-pulse structure. 
We hypothesize that the states observed here correspond to different snaking branch solutions reported in the bifurcation analysis of switching waves~\cite{Parra-Rivas:16} that become connected with the pump detuning. The number and shape of the low intensity oscillations in the dark-pulses vary continuously with the pump tuning, from one comb state to another. Similarly, the comb power in Fig.~\ref{fig:Exp_Switching}a changes gradually from one step to another, as opposed to the abrupt jumps observed in dissipative Kerr solitons~\cite{Guo2016}. 

After accessing a comb state in the forward tuning, the switching transition is reversed by tuning the pump backwards (yellow curve in Fig.~\ref{fig:Exp_Switching}a). The comb power drops and shows a step-like pattern similar to the forward tuning, but in the reverse direction. 
Thus, the low intensity oscillations in the dark-pulse vanish one by one, until what appears to be a single {\color{black}gray soliton state~\cite{Weiner:gray1,Weiner:gray2,KIVSHAR98} is accessed (state F in Fig.~\ref{fig:Exp_Switching}d). However, strictly speaking it is not a gray soliton in the sense of the dissipationless nonlinear Schr\"{o}dinger equation} as the temporal phase is not an odd function of time due to the periodic boundary conditions of the cavity. 
Switching occurs over a broader detuning range 
in the backward pump tuning (state C to F) compared to forward tuning (State A to C), giving access to more comb states. The comb power in the forward and backward pump tuning shows a hysteresis behavior, which is similar to what has been observed for dissipative Kerr solitons in the anomalous dispersion regime~\cite{Guo2016}.

We find an excellent agreement between the measured and simulated comb spectra, indicating that 
by just measuring the transmission spectrum and retrieving the parameters of the interacting modes, one can predict the comb dynamics starting from a continuous-wave pump by using two linearly coupled equations. The discrepancy between simulations and experiments for longer wavelengths in Fig.~\ref{fig:Exp_Switching}d might be due to a second mode coupling around 1590 nm. The measurements display an asymmetry in the comb spectra, which had also been observed in previously reported dark-pulse Kerr combs~\cite{Attila2018}. Our simulations capture naturally this comb asymmetry even though the third-order dispersion has not been included, clearly indicating that the asymmetry is caused by the linear coupling between the two transverse modes. 
\subsection{Hot-cavity spectroscopy of dark-pulse Kerr combs.}
\begin{figure}
\centering
\includegraphics[width=1\linewidth]{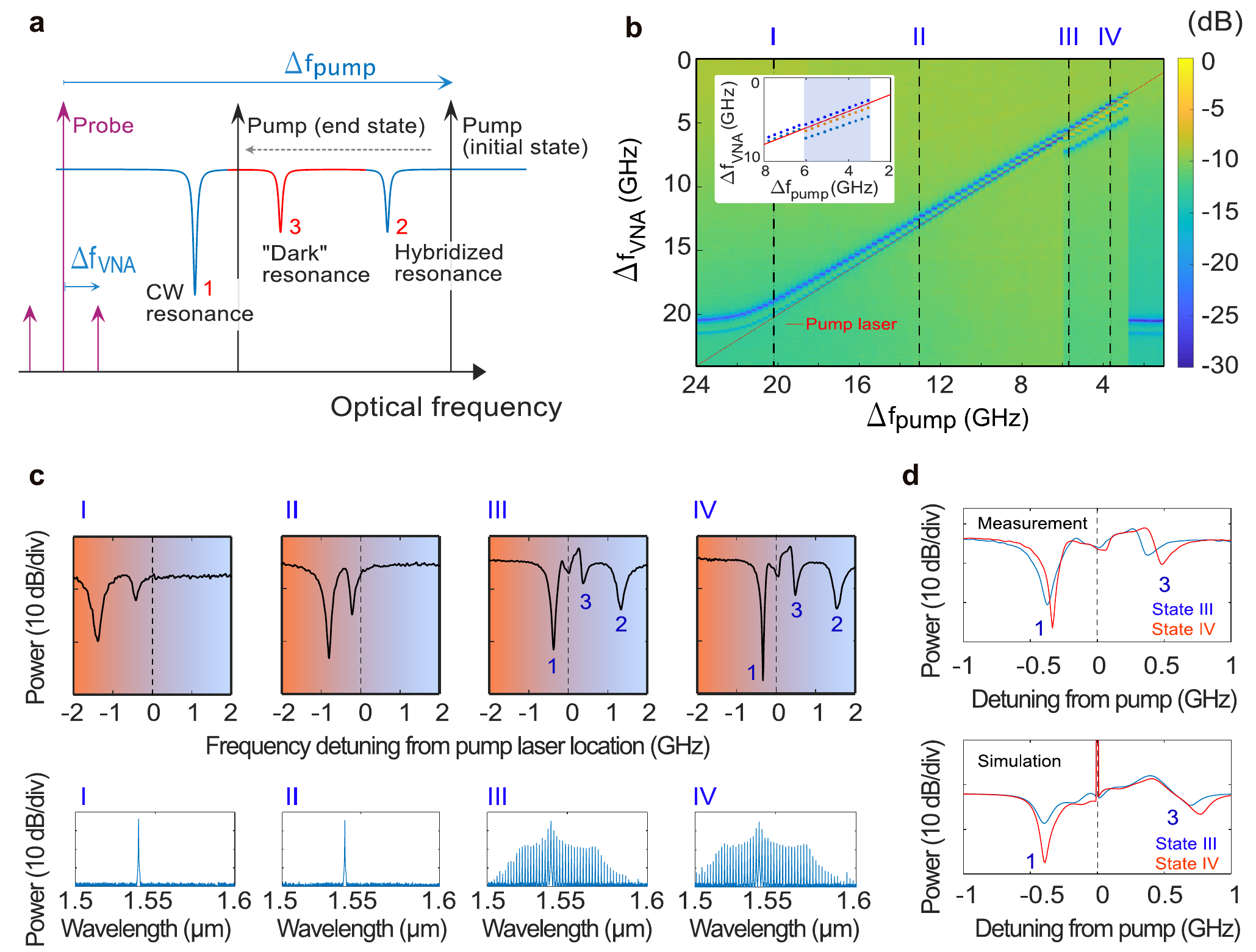}
\caption{Hot-cavity spectroscopy of dark-pulse Kerr combs. 
\textbf{a}, Schematic diagram of the method used for the measurement of the system response. 
\textbf{b}, System response as the pump is tuned into resonance. The inset shows a zoomed-in view, where the existence of dark-pulses is highlighted. The appearance of an extra resonance can be clearly observed in this regime. 
\textbf{c}, VNA traces and corresponding comb spectra for various pump detunings indicated by dashed lines in \textbf{b}. Note that the origin of the frequency axis here is the pump laser, thus providing a direct indication of the location of the resonances and effective pump-laser detuning.  
\textbf{d}, Measured and simulated VNA traces associated with the main mode for pump detunings III and IV in \textbf{b}. The experimental traces are the zoomed-in view of the states III and IV in \textbf{c}.
}
\label{fig:VNA}
\end{figure}

To get a better understanding of the dynamics of dark-pulse Kerr combs, we look into the system's response upon the formation of the dark-pulse. The formation of a dark-pulse would break the time invariance of the system, making it impossible to describe the resonator in terms of a linear transfer function. Instead, we measure the system's response with the aid of an external (probe) laser, as sketched in Fig.~\ref{fig:VNA}a. A probe laser with a fixed frequency, far detuned from the cavity resonances, is weakly modulated with an external modulator driven by a tunable radio-frequency signal, allowing to retrieve the system's response as the pump laser is tuned into resonance. The benefit of using this scheme instead of modulating the pump itself is that it solves the ambiguity in the pump location with respect to the resonances of the coupled modes. The probe sideband will be affected by the presence of resonances in the cavity and nonlinear distortions caused by the pump and the dark-pulse. A vector network analyzer (VNA) measures the magnitude of the radio-frequency beat between the sideband and probe laser as the sideband is swept in frequency. Further details of the measurement scheme are described in the Methods section. This process is repeated for various pump frequency detunings, corresponding to different states, and the recorded system's responses are displayed in Fig.~\ref{fig:VNA}b. The parameter $\Delta \text{f}_\text{pump}$ determines the location of the pump laser, while $\Delta \text{f}_\text{VNA}$ is the detuning of the sideband, both measured relative to the probe laser. The pump laser approaches the hybridized resonances from the blue side, {\color{black}meaning that $\Delta \text{f}_\text{pump}$ decreases as the laser is tuned closer to the resonances.} At the initial stages prior to the formation of the dark-pulse Kerr comb, the system's response is affected by thermal dynamics. {\color{black}Consequently, the resonances of the hybridized modes are strongly red shifted, which decreases their $\Delta \text{f}_\text{VNA}$} (states I and II in Figs.~\ref{fig:VNA}b,c). By further tuning in the pump, a mode-locked dark-pulse comb emerges suddenly (state III) as soon as the pump crosses the first hybridized resonance labeled as ``2'' in Fig.~\ref{fig:VNA}a. The generation of a dark-pulse Kerr comb is associated with the emergence of a third resonance (labeled ``3'' in Fig.~\ref{fig:VNA}a and visible in the states III and IV in Figs.~\ref{fig:VNA}b,c). This feature has a striking similarity to observations made for DKS~\cite{Guo2016}. In this stage, the pump frequency is effectively red-detuned with respect to this resonance. {\color{black}Once the pump crosses the hybridized resonance ``2'', this resonance cools down and rapidly moves back towards its original location (higher frequency). Our simulations, which will be explained in the following (see Fig.~\ref{fig:VNA}d), confirm that the newly generated resonance ``3'' is the resonance  located closer to the pump, while the resonance further away from the pump is the hybridized resonance ``2''.} Further moving the pump to the red side causes the laser to cross the two hybridized resonances, associated with a loss of the comb and a cooling down of the system (see Fig.~\ref{fig:VNA}b). The system's response in this case is similar to that of the first stage, with two Lorentzian shape resonances associated with the hybridized modes. 

We explain the appearance of the third resonance as follows. Dark-pulses are intermediate solutions between the upper and lower CW steady state solutions of the Lugiato-Lefever equation~\cite{Lugiato,Coen:13,HAELTERMAN:92,Godey}. Given the intensity dependence of the Kerr effect, the high intensity CW background and the low intensity oscillations in the dark-pulse induce different nonlinear Kerr phase shifts on the cavity resonances, but most notably on the main mode. In particular, the high-power level shifts the cavity resonance to the red side of the pump, while the low-power level induces a smaller shift on the cavity resonance and creates a resonance on the blue side of the pump. This behavior is analogous to the appearance of the soliton resonance in the system transfer function of anomalous dispersion microresonators~\cite{Guo2016}. In dark-pulse Kerr combs, a subtle yet important difference is that the CW background corresponds to the high-power level, so its resonance appears on the red side of the pump. Meanwhile, the low intensity oscillations at the center of the pulse, which are associated with the generation of the dark-pulse, generate a new resonance on the blue side of the pump labeled as ``dark resonance" (Fig.~\ref{fig:VNA}a). Hence, in contrast to dissipative Kerr solitons, where the CW background is weak and the soliton has a high-power level, the pump laser remains on the effectively blue-detuned side for dark-pulse Kerr combs with respect to the CW background resonance (labeled ``1'' in Fig.~\ref{fig:VNA}a). The observations made here are consistent with previous studies {\color{black}based on modulation of the resonator using a microheater~\cite{Xue2015}}. The emergence of an extra resonance is a unique property of stationary solitonic states in Kerr microresonators that had not been previously demonstrated for dark-pulse Kerr combs. We also measured the system's response for another microresonator chip operating in the normal dispersion regime and observed a similar behavior, validating the generality of our observations (Supplementary Note 1).

We support the explanation of the
VNA response above with numerical simulations. For simplicity, the simulation assumes only a single transverse mode in the cavity, with the initial
intracavity field taken as a square dark-pulse (see Methods). {\color{black}This is a reasonable assumption given that the appearance of the new resonance in the system's response arises from the two power levels present in the intracavity waveform (related to the main mode), and does not depend on the auxiliary mode nor the linear coupling between the modes.}
Thus, only the resonances related to the main mode are considered. For comparison, the measured and simulated system responses at two different pump detunings in the dark-pulse regime are shown in Fig.~\ref{fig:VNA}d. The appearance of an extra resonance is evident in the simulations too. Both measured and simulated results indicate that the depth of the CW resonance increases with the red tuning of the pump. Moreover, the frequency of the CW resonance remains almost fixed, while that of the dark resonance varies with the pump detuning. {\color{black}The power variations of the VNA traces around the pump frequency, observed in both experiments and simulations, are related to the nonlinear effects induced on the sideband; an aspect that has also been observed in other experiments using other microresonator platforms~\cite{Ghalanos:19}.} Note that the switching between dark-pulse comb states is not associated with sharp changes in the system's response. The reason is that switching in dark-pulses changes the number of oscillations, which have a low intensity and do not introduce a significant energy change in the cavity. 
\section*{Discussion}
The physics of dark-pulse Kerr comb generation and its switching dynamics are investigated in this work, both experimentally and numerically. Deterministic switching between dark-pulse comb states is observed, in which the number of low intensity oscillations at the center of the corresponding dark-pulses can either increase or decrease, one at a time. 
Moreover, we measure the system's response as the pump laser is tuned into resonance and discover that the formation of a dark-pulse Kerr comb is associated with the emergence of an extra resonance. 
This is due to the combination of nonlinearity and bi-stability in the cavity.  
The revealed multi-resonance dynamics is a distinctive property of soliton states in Kerr microresonators and confirms the switching behavior of the dark-pulse combs in a new way.
Furthermore, by using an external probe to measure the system's response, we could disentangle the different resonances present in the system, clearly indicating that for dark-pulse states, the pump laser lies in the effectively blue-detuned region 
{\color{black}with respect to the CW background resonance}, in sharp contrast to dissipative Kerr solitons in anomalous dispersion microresonators. These results shed light into the formation of Kerr combs in normal dispersion microresonators and pave the way for the generation of reproducible chip-scale comb sources with high power conversion efficiency. 

\begin{methods}
\subsection{Microresonator operation.}
The microresonator chip is placed on a piezo-controlled positioning stage which is temperature controlled with a standard laser temperature controller at 18$~^{\circ}$C, limiting the variations to less than 0.01$~^{\circ}$C. This allows stable comb operation over several hours. The pump power is amplified in an erbium-doped fiber amplifier and optically filtered to remove the amplified spontaneous emission noise far away from the pump. This increases the signal quality of the generated comb lines. Then, it is coupled into the microresonator using a lensed fiber. The off-chip power is 25.6 dBm. 
At high pump powers, the coupling losses between the fiber and chip are estimated to be 4-5 dB per facet. The simulations have a good agreement with the experimental measurements for an input pump power of 150 mW (21.8 dBm). The coupling between the bus waveguide and the ring corresponds to the 300 nm wide gap between them.  

\subsection{System's response measurement with the VNA.}
A tunable pump laser is amplified and filtered before coupling into the microresonator. The pump is tuned across the cavity resonance from the blue side, over a 22.88 GHz bandwidth ($\Delta \text{f}_\text{pump}$ varying from 23.94 to 1.053~GHz) in 81 steps. Meanwhile, a red-detuned probe laser that is fixed in frequency is weakly modulated using an external, dual-sideband electro-optic intensity modulator driven by an external radio-frequency (RF) source. One of the two generated sidebands is scanned across the cavity resonances of the two interacting modes, by sweeping the RF source from 10 MHz to 24 GHz in 10 MHz steps. A vector network analyzer performs the RF sweeping and measures the magnitude of the beat between the sideband and the probe laser. This is done for each pump detuning, leading to the results presented in Fig.~\ref{fig:VNA}b. The two lasers are not locked to each other, so the measurement of the system's response has a frequency resolution in the order of a few MHz, given by the relative drift between the probe and pump lasers. Since the pump frequency varies in every step, the beat note between the pump and sideband is also used to find the location of the pump. We monitor the comb states generated in the microresonator using an optical spectrum analyzer, while the power in the generated comb lines is measured with an oscilloscope, after suppressing the remaining pump line with an optical filter.

\subsection{Dark-pulse comb numerical simulations.}
The simulations are based on a modified Ikeda map~\cite{IKEDA,Torres-Company:14,Hansson:15} that considers the linear mode interaction in the microresonator. Each round trip has two steps, one is the coupling between the pump in the bus waveguide and the ring and the other is the light propagation in the microresonator.
The coupling between the pump to the modes in the resonator is found through coupled mode theory~\cite{Haus,Yariv}, assuming a 3$\times$3 lossless directional coupler. It can be expressed as 
\begin{equation}\renewcommand{\arraystretch}{1.}
\left[ \begin{array}{c} A_{out} \\ A_1^{(k+1)} \\ A_2^{(k+1)} \end{array} \right] = 
\renewcommand{\arraystretch}{1.7}
\begin{bmatrix} 
\sqrt{1-\theta_1-\theta_2} & 
-\sqrt{\theta_1} & 
-\sqrt{\theta_2}\\  

\sqrt{\theta_1} & \dfrac{\theta_2+\theta_1\sqrt{1-\theta_1-\theta_2}}{\theta_1+\theta_2} & 
\dfrac{\sqrt{\theta_1\theta_2}}{\theta_1+\theta_2}(\sqrt{1-\theta_1-\theta_2}-1)\\ 

\sqrt{\theta_2} &
\dfrac{\sqrt{\theta_1\theta_2}}{\theta_1+\theta_2}(\sqrt{1-\theta_1-\theta_2}-1) & 
\dfrac{\theta_1+\theta_2\sqrt{1-\theta_1-\theta_2}}{\theta_1+\theta_2}
\end{bmatrix} 
\renewcommand{\arraystretch}{1.}
\left[ \begin{array}{c} A_{in} \\ A_1^{(k)} \\ A_2^{(k)} \end{array} \right]
\end{equation}
where $A_{in}$ and $A_{out}$ are the input pump and throughput field of the microresonator. $A_{m}^{(k)}$ is the intracavity field of mode $m$ at roundtrip $k$ in the cavity and $\theta_m$ is the coupling coefficient between the bus and the microring for mode $m$. 
The evolution of the fields in the resonator is modeled using the nonlinear Schr\"{o}dinger equation in multimode waveguides~\cite{Poletti:08,Mafi:12}. The propagation of mode $m$ in every round-trip of the microresonator is given by 
\begin{equation}
\begin{split}
\dfrac{\partial A_{m}}{\partial z}&=
\dfrac{-\alpha_m}{2}A_{m}+i\beta^{(m)}_0 A_{m}-\beta^{(m)}_1 \dfrac{\partial A_{m}}{\partial t}
-i\dfrac{\beta^{(m)}_2}{2} \dfrac{\partial^2 A_{m}}{\partial t^2}
\\&
+i\gamma_{m}~|A_{m}|^2 A_{m}
+i\kappa_{mn} A_{n\neq m},
\end{split}
\label{eq:NLSE}
\end{equation}
where 
$\alpha_{m}$ is the propagation loss, $\beta^{(m)}_1$ is the inverse group velocity, $\beta^{(m)}_2$ is the group velocity dispersion, and $\gamma_m$ is the nonlinear coefficient of mode $m$, while $\kappa_{mn}$ is the mode coupling strength between modes $m$ and $n$. The resonator length is $L$ and the linear phase shift of the field is $\beta^{(m)}_0 L=-\delta^{(m)}_0$, where $\delta^{(m)}_0$ is the pump detuning from the cold-cavity resonance of each transverse mode closest to the pump frequency $\omega_p$. It can be expressed as $\delta^{(m)}_0=[\beta^{(1)} (\omega_0)-\beta^{(m)} (\omega_p)] L$, where $\beta^{(m)} (\omega)$ is the propagation constant of mode $m$ at frequency $\omega$ and $\omega_0$ denotes the pumped resonance frequency of the main mode. {\color{black}
Note that the nonlinear mode coupling~\cite{Menyuk} is not included in our simulations using Eq.~\ref{eq:NLSE}, assuming that it is negligible. The excellent agreement between the experimental and simulated comb spectra confirm that the linear mode coupling is the dominant cause leading to the generation of the dark-pulse comb.}

In each round trip a CW pump together with quantum noise consisting of one photon per spectral bin with random phase~\cite{Torres-Company:14} is coupled to the ring. 
Unlike previously reported models that start the simulations with an initial intracavity square dark-pulse, the initial intracavity field here is just the quantum noise in the resonator. The closest resonance to the avoided crossing is pumped from the blue side and the detuning is dynamically changed to emulate the experiments. The propagation in the ring is carried out using the split-step Fourier method. For the main mode, $\alpha_1$ corresponds to 0.1 dB/cm, 
$\theta_1=0.004$, the initial pump detuning is $\delta^{(1)}_0=-0.001$ rad, $\beta^{(1)}_2=139~\rm{ps^2/km}$ and $\text{FSR}_1=229.08~\rm{GHz}$. For the auxiliary mode, $\alpha_2$ corresponds to 0.3 dB/cm, $\theta_2=0.01$, $\delta^{(2)}_0=-0.0033$ rad, $\beta^{(2)}_2=1.8~\rm{ps^2/km}$ and $\text{FSR}_2=221.45~\rm{GHz}$. {\color{black}Note that the main and auxiliary modes and their corresponding detunings mentioned here are associated with the uncoupled and cold-cavity system, while the measurements correspond to the hot-cavity hybridized modes.} The nonlinear coefficients are $\gamma_1=0.89~\rm{m^{-1}W^{-1}}$ and $\gamma_2=0.44~\rm{m^{-1}W^{-1}}$, and the ring length is $\text{L}=2\pi \times 100~\mu \rm{m}$.
The linear coupling between the two modes is $\kappa_{12}=22.7~\rm{m^{-1}}$, calculated from the measurements in Fig.~\ref{fig:ring}d. The pump detuning is varied linearly in a dynamic manner, such that the final detuning of the main mode is $\delta^{(1)}_0=0.02$ rad after 750 ns. 
After the field inside the cavity has stabilized and converged to a steady state, the results are analyzed.

\subsection{Simulation of system's VNA response.}

The simulations are performed based on the propagation of only a single mode in the cavity, using the Ikeda map~\cite{IKEDA,Torres-Company:14,Hansson:15}. The initial intracavity field is a square dark-pulse, where the amplitude
and phase of the top (bottom) of the pulse are equal to the upper branch (lower branch) steady-state values~\cite{Xue2015}. The considered ring parameters and pump power are similar to the main mode values mentioned in the previous section. The pump is fixed in frequency and a weak (-40~dBm) probe 
is swept across the resonance in 10 MHz steps. In each step, 
after simulating the output spectrum of the microresonator, the power of the probe frequency component is calculated.  
The comparison between this power and the probe power at the input is the system's transfer function. It corresponds to the beat note between the sideband and the resonances, measured with the VNA in the experiment. The considered {\color{black} cold-cavity} detunings are $\delta^{(1)}_0=0.0242$~rad and $\delta^{(1)}_0=0.0248$~rad, which correspond to the comb states III and IV in Fig.~\ref{fig:VNA}, respectively.

\end{methods}

\bibliographystyle{naturemag}
\bibliography{Main}

\begin{thebibliography}{10}
\bibitem{note1}
\bibinfo{note}{Note that we use the standard terminology in
  nonlinear~\cite{Agrawal} and ultrafast optics~\cite{Weiner}, where anomalous
  dispersion refers to a negative group velocity dispersion coefficient,
  whereby the blue components in the microresonator travel faster than the red
  ones, leading to an increasing free spectral range (FSR) with optical
  frequency.}
\end{thebibliography}

\begin{addendum}
 \item This work was supported in part by the European Research Council (ERC
CoG, grant agreement 771410 DarkComb), Swedish Research Council (VR), the National Science Foundation (NSF) (ECCS-150957), DARPA (W31P40-13-1-001.8) and AFOSR (FA9550-15-1-0211).
 \item[Competing Interests] The authors declare that they have no
competing financial interests.
 \item[Correspondence] Correspondence and requests for materials
should be addressed to V.T-C.~(email: torresv@chalmers.se).
\end{addendum}

\end{document}


\maketitle

\section*{Supplementary Note 1 : Hot-cavity spectroscopy of dark-pulse Kerr combs.}

\begin{figure}[h]
\centering
\includegraphics[width=1\linewidth]{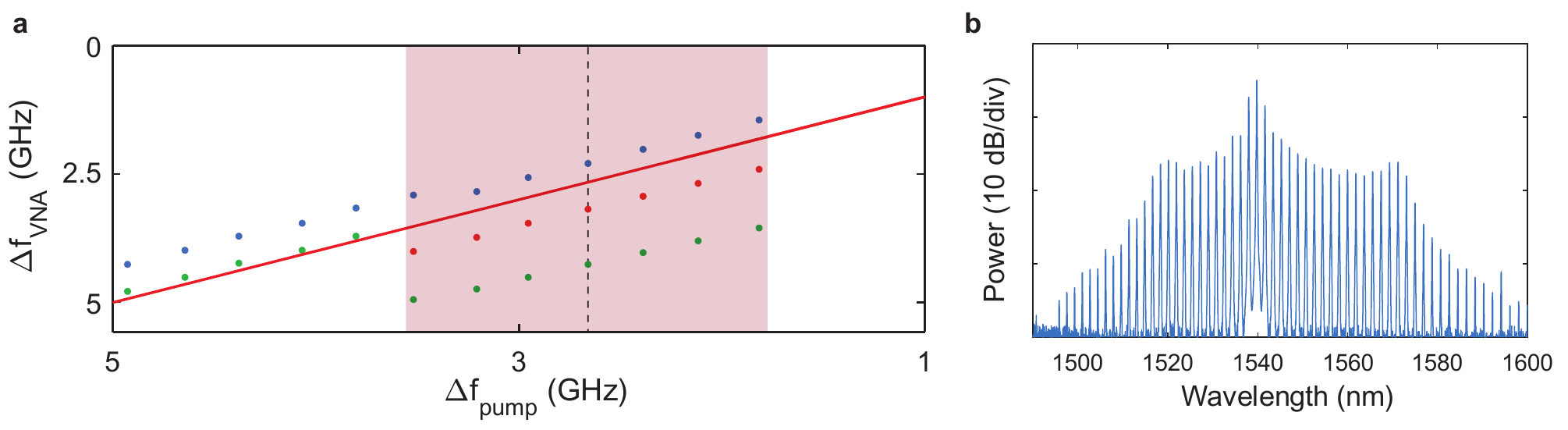}
\caption{Hot-cavity spectroscopy of dark-pulse Kerr combs. 
\textbf{a}, System response as the pump is tuned into resonance from the blue side. The appearance of a third resonance can be clearly observed in the dark-pulse regime (highlighted region) for a different silicon nitride resonator than the one used in the main manuscript. 
\textbf{b}, Dark-pulse Kerr comb spectrum measured at the pump detuning indicated by a dashed line in \textbf{a}.
}
\label{fig:Supp}
\end{figure}

To show the generality of our observations, we also measured the system's response in a second silicon nitride microresonator chip. The microresonator has nominally the same dimensions as the chip used in our main experiments, except for the gap between the ring and the drop-port. The measured results are shown in Fig.~\ref{fig:Supp}. Similar to the observations presented in the main manuscript, the formation of a dark-pulse Kerr comb in this microresonator is also clearly accompanied by the emergence of a new resonance. 
